\documentclass[prb,a4paper,twocolumn,showpacs,preprintnumbers,amsmath,amssymb,floatfix]{revtex4}

\usepackage{graphicx}
\usepackage{dcolumn}
\usepackage{bm}

\newcommand{\Slowlevel}[1]{(TMTSF)$_2$#1}
\newcommand{\SX}{\Slowlevel{$X$}}
\newcommand{\NO}{\Slowlevel{NO$_3$}}
\newcommand{\PF}{\Slowlevel{PF$_6$}}
\newcommand{\ReO}{\Slowlevel{ReO$_4$}}
\newcommand{\ClO}{\Slowlevel{ClO$_4$}}
\newcommand{\ET}{$\alpha-$(ET)$_2$KHg(SCN)$_4$}

\newcommand{\myreffig}[1]{Fig.\ \ref{#1}}

\DeclareMathOperator{\sech}{sech}
\newcommand{\mdeg}{^\circ}
\newcommand{\mTSDW}{T_C}

\newcommand{\mv}[1]{\mathbf{#1}}               

\newcommand{\mva}{\mv{a}}  \newcommand{\va}{$\mva$}
\newcommand{\mvb}{\mv{b}}  \newcommand{\vb}{$\mvb$}
\newcommand{\mvbp}{\mv{b'}}  
\newcommand{\mvc}{\mv{c^\star}}   \newcommand{\vc}{$\mvc$}
\newcommand{\mvk}{\mv{k}}   
\newcommand{\mvB}{\mv{B}}  \newcommand{\vB}{$\mvB$}

\newcommand{\plane}[2]{#1\text{--}\,#2}
\newcommand{\mbcplane}{\plane{\mvbp}{\mvc}}
\newcommand{\bcplane}{$\mbcplane$}

\begin{document}


\title{Unconventional spin density wave in Bechgaard salt \NO}

\author{Mario Basleti\'{c}}
    \affiliation{Department of Physics, Faculty of Science, P.O.Box 331, HR-10002 Zagreb, Croatia}
    \email{basletic@phy.hr}

\author{Bojana Korin-Hamzi\'{c}}
    \affiliation{Institute of Physics, P.O.Box 304, HR-10001 Zagreb, Croatia}

\author{Kazumi Maki}
    \affiliation{Department of Physics and Astronomy, University of Southern California, Los
Angeles CA 90089-0484, USA\\ and\\ Max Planck Institute for the
Physics of Complex Systems, N\"{o}thnitzer Str.\ 38, D-01187,
Dresden, Germany}

\author{Silvia Tomi\'{c}}
    \affiliation{Institute of Physics, {P}.O.Box 304, HR-10001 Zagreb, Croatia}
\date{\today}

\begin{abstract}
Among many Bechgaard salts, \NO\ exhibits very anomalous low temperature properties. Unlike conventional spin density wave (SDW), \NO\ undergoes the SDW transition at $\mTSDW\approx9.5\,$K and the low temperature quasiparticle excitations are gapless. Also, it is known that \NO\ does not exhibit superconductivity even under pressure, while FISDW is found in \NO\ only for $P=8.5\,$kbar and $B>20\,$T. Here we shall show that both the angle dependent magnetoresistance data and the nonlinear Hall resistance of \NO\ at ambient pressure are interpreted satisfactory in terms of
unconventional spin density wave (USDW). Based on these facts, we propose a new phase diagram for Bechgaards salts.
\end{abstract}

\pacs{74.70.Kn 72.15.Gd 75.30.Fv 71.70.Di}

\maketitle

\section{Introduction}
\SX\ are quasi one dimensional molecular conductors, known as Bechgaard salts, where TMTSF denotes tetramethyltetraselenafulvalene and $X$ is an inorganic anion with various possible symmetries: spherical (octahedral) ($X$$=$PF$_6$, AsF$_6$\ldots), tetrahedral ($X$$=$ClO$_4$, ReO$_4$\ldots), or triangular (NO$_3$). The well known are their very complex (pressure, magnetic field, temperature) phase diagrams with a variety of electronic ground states: (conventional) spin density wave (SDW), field induced SDW (FISDW), superconductivity (triplet, unconventional), unconventional spin density wave (USDW).\cite{IshiguroBook98,MoriJPSJ06,Schegolev96,LeePRL02,JooEPL05,LeeJPSJ06,KangSM03,DoraEPL04} The observation of superconductivity in the \SX\ series requires the use of high pressure, with the exception of \ClO, which is superconducting under ambient pressure, and \NO, which never becomes a superconductor even under pressure.

NO$_3$ anions are in an orientational disorder at ambient pressure and for $T>45\,$K. The anion ordering (AO) transition takes place at $T_{\text{AO}}\approx45\,$K, with wave vector $\mathbf{q}=(1/2,0,0)$. Contrary to AO in most other salts, $\mathbf{q}$ has the nonzero component parallel to the most conducting direction. The SDW state develops below $\mTSDW\approx9.5\,$K. From the resistivity data very small activation energy was obtained, of order of $10^{-3}\,$eV, but the curvature of the $\log R$ vs.\ $1/T$ plot indicated that the ground state should be considered as semimetalic, rather than semiconducting.\cite{TomicPRL89}

The phase diagram of Bechgaard salts under pressure is interpreted in terms of the standard model, where the approximate nesting of the quasi-one dimensional Fermi surface (i.e.\ the imperfect nesting), and the repulsive Coulomb interaction between electrons are the crucial ingredients.\cite{YamajiSM86,VirosztekPRB86,PoilblancJPC86} The applied pressure increases the 2-dimensionality of Bechgaard salts through the increase of the imperfect nesting term.\cite{IshiguroBook98} However, the standard model does not yet describe neither the triplet\cite{LeePRL02,LeeJPSJ06} superconductivity nor USDW.

UDW is a density wave, whose gap function depends on the wavevector, vanishes on certain subsets of the Fermi surface, allowing for low energy excitations. The average of the gap function over the Fermi surface is zero, causing the lack of periodic modulation of the charge and/or spin density. As noted by Nersesyan et al.\ (Refs.\ \onlinecite{NersesyanJLTP89} and \onlinecite{NersesyanJPCM91}), the quasiparticle spectrum in UDW is quantized in a magnetic field. This Landau quantization gives rise to the spectacular angle dependent magnetoresistance (ADMR) and giant Nernst effect.\cite{DoraMPL04,MakiCondMat0603806} As we shall see later both the angle dependent magnetoresistance\cite{BiskupPRB94} and the nonlinear Hall resistance\cite{BasleticSSC96} of \NO\ are described nicely in terms of USDW. We note that an earlier attempt to describe the magnetoresistance of \NO\ in terms of conventional SDW with a {\it large imperfect nesting} might not be the most appropriate model,\cite{BiskupPRB93} since it cannot describe the details of the resistance quantitatively. We also propose the revision of the generally accepted phase diagram, taking into account the identification of USDW state in several Bechgaard compounds.\cite{KangSM03,DoraEPL04,BasleticPRB02}

\section{Identification of USDW}

Here we summarize briefly what is known about unconventional density wave.\cite{DoraMPL04,MakiCondMat0603806} The unconventional density
wave is a kind of density wave, where the quasiparticle energy gap
vanishes along lines on the Fermi surface. In the present instance
we can assume $\Delta(\mvk) \sim \cos\mvb\mvk$ or $\Delta(\mvk) \sim
\sin\mvb\mvk$ as in earlier analysis of UCDW in
\ET.\cite{DoraEPL02,MakiPRL03} Then the quasiparticle Green function
is given by
\begin{equation}\label{eq:1}
G^{-1}(\mvk,\omega) = \omega- \eta(\mvk) -\xi(\mvk)\rho_3 -
\Delta(\mvk)\rho_1
\end{equation}
where $\rho_i$'s are the Pauli matrices and $G$ operates on the
Nambu spinor space.\cite{NambuPR60} The quasiparticle energy in UDW
is formed from the pole of $G(\mvk,\omega)$ as
\begin{equation}\label{eq:2}
\omega = \eta(\mvk) \pm \sqrt{\xi^2(\mvk)+\Delta^2(\mvk)}
\end{equation}
where $\xi(\mvk)$ is the kinetic energy of electrons measured from the Fermi energy in the normal state, $\xi(\mvk) \simeq v(k_{\mva}-k_F)$, $\eta(\mvk)$ is the
imperfect nesting term and $\Delta(\mvk)=\Delta\cos\mvb\mvk$. Here, $v$ denotes the Fermi velocity in the chain (\va) direction, $\Delta$ is the order parameter for unconventional SDW, and \vb\ is the lattice constant.

Then in a magnetic field \vB\ in the \bcplane\ plane with angle
$\theta$ from the \vc\ axis the quasiparticle energy changes into
\begin{equation}\label{eq:3}
E_n^{\pm} = \pm\sqrt{2neB|\cos\theta|vb\Delta}
\end{equation}
with $n=0,1,2\ldots$. Here we have neglected $\eta(\mvk)$ for
simplicity. Also in the following we assume $b=7.567\,$\AA\ and
$v=3\times10^5\,$m/s.\cite{BiskupPRB93} Eq.\ (\ref{eq:3}) is the
consequence of the Landau quantization of the quasiparticle spectrum
in UDW, or the {\it Nersesyan effect}.\cite{NersesyanJPCM91,NersesyanJLTP89}

Then the conductivity tensor is constructed as
\begin{equation}\label{eq:sxx}
\sigma_{xx}=\sigma_1(1+2C_1\sech^2(x_1/2) + \cdots)
\end{equation}
\begin{equation}\label{eq:syy}
\sigma_{yy}=\sigma_2(1+2C_2\sech^2(x_1/2) + \cdots)
\end{equation}
\begin{equation}\label{eq:sxy}
\sigma_{xy}=\sigma_3 n(T,B) B |\cos\theta|
\end{equation}
with
\begin{equation}\label{eq:n}
n(T,B)=n_0\left[1+2(1-\tanh(x_1/2)) + \cdots\right]
\end{equation}
where $x_1=E_1/k_BT$ and we have assumed that $x_1\gg1$. Also, we
have assumed that $\sigma_1$, $\sigma_2$, $\sigma_3$, $C_1$ and
$C_2$ are weakly dependent on $T$ and $B$. Then from Eqs.\
(\ref{eq:sxx})--(\ref{eq:sxy}) we can construct the resistivity
tensor as
\begin{equation}\label{eq:Rxx}
R_{xx}(B,\theta)=\frac{R_0}{1-D_1\tanh^2(x_1/2)}\,
\end{equation}
\begin{equation}\label{eq:Rxy}
R_{xy}(B,T)=\frac{D_2B}{\displaystyle\frac{n(B,T)}{n(0,T)}B^2
    +D_3 \frac{n(0,T)}{n(B,T)}\sigma_{xx}\sigma_{xy}
        }.
\end{equation}

\begin{figure}
\includegraphics*[width=8.6cm]{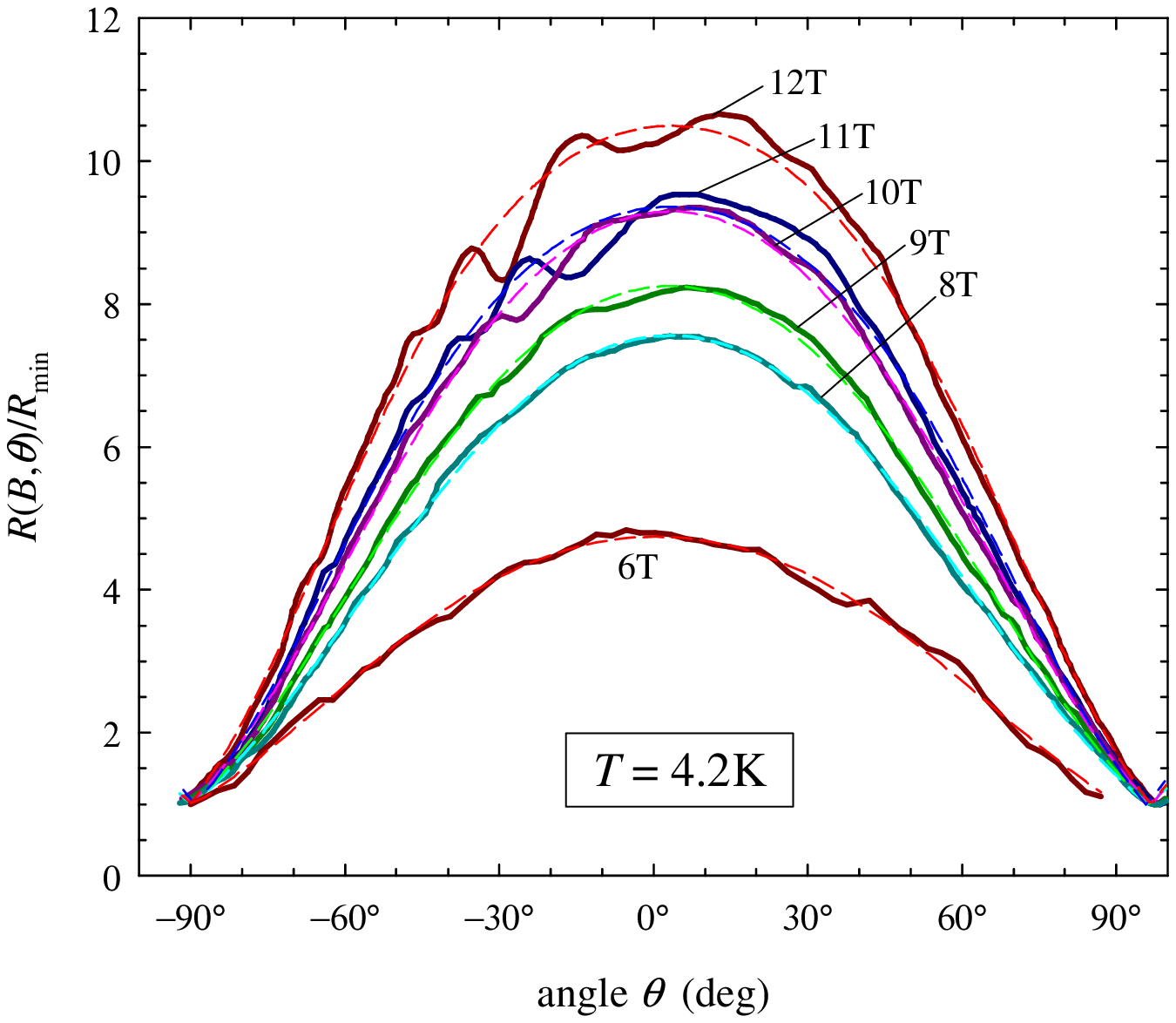}
\caption{\label{fig:admr}(Color online) The angular dependence of the normalized
resistance $R(B,\theta)/R_0$ at $T = 4.2\,$K (full lines:
experimental data; dashed lines: fits to the theory). Magnetic field
is rotated in \bcplane\ plane, and $\theta=0\mdeg$ corresponds to
$\mv{B}\|\mvc$. Data are from Ref.\ \onlinecite{BiskupPRB94}
($B\geq8\,$T) and Ref.\ \onlinecite{BasleticEPL93} ($B=6\,$T).}
\end{figure}

In \myreffig{fig:admr} we show our fitting of the angle dependent
magnetoresistance data for \NO\ at $T=4.2\,$K for a variety of
magnetic field.\cite{BiskupPRB94,BasleticEPL93}
From this fitting we obtain USDW order parameter $\Delta=6.3\,$K and $D_1=0.93$. As is
readily seen the fitting is excellent except for the bumpy
structures. These should come from the imperfect nesting term as
discussed in Refs.\ \onlinecite{DoraEPL04}, \onlinecite{DoraEPL02}
and \onlinecite{MakiPRL03}. Also we note $D_1\approx 2C_1/(1+2C_1)$
indicating that $C_1=7.1$; therefore $\sigma_{xx}$ is dominated by the $n=1$
excitations.

\begin{figure}
\includegraphics*[width=8.6cm]{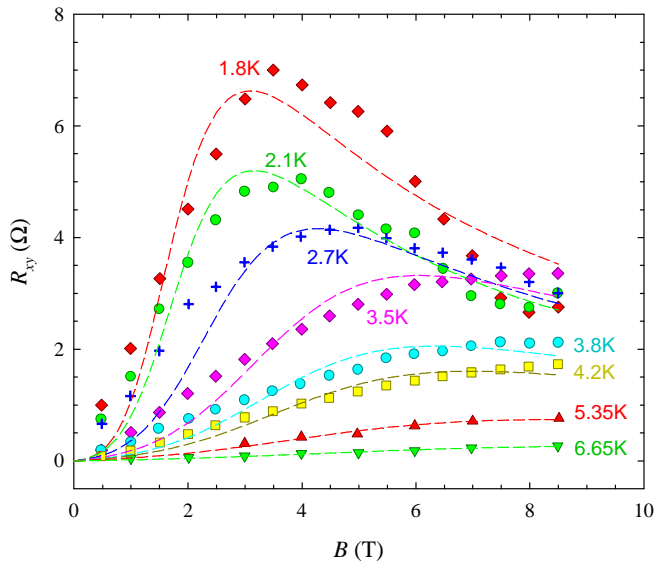}
\caption{\label{fig:Rxy}(Color online) The magnetic field $B$ dependence of Hall
resistance $R_{xy}$ at several temperatures (points: experimental
data; dashed lines: fit to the theory). Data are from Ref.\
\onlinecite{BasleticSSC96}.}
\end{figure}

\begin{figure}
\includegraphics*[width=8.6cm]{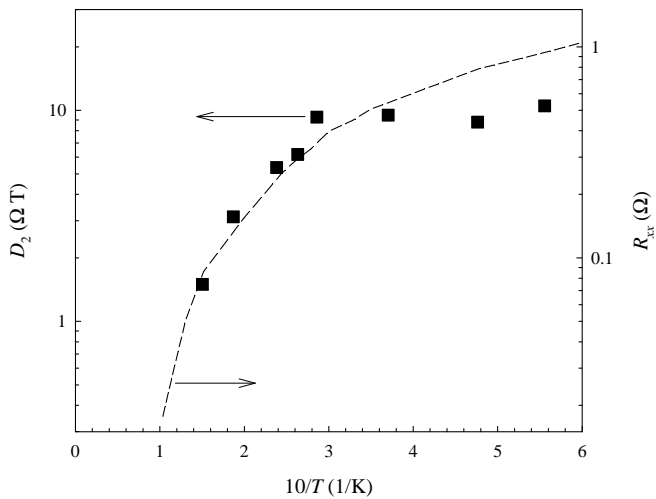}
\caption{\label{fig:fitparams}The temperature dependence of the fitting parameters $D_2$ ($\blacksquare$) -- see text. Dashed line shows temperature dependence of the resistance $R_{xx}$ of \NO\ for $B=0$.}
\end{figure}

In \myreffig{fig:Rxy} we show $R_{xy}$ fitted by Eq.\
(\ref{eq:Rxy}); again we obtain reasonable fitting with $D_3\simeq 80\,\Omega$T.
Figure \ref{fig:fitparams} shows temperature dependence of the parameter $D_2$, along with
 temperature dependence of the resistance $R_{xx}$. There appears to be a slight change of the parameter $D_2$ across $T^\star\approx T_C/3=3\,$K: for $T \gtrsim 3\,$K it follows temperature dependence of resistance $R_{xx}$, while for lower temperature it is nearly constant. It signals the possible occurrence of yet another phase transition at $3\,$K -- as in \PF, in agreement with several other suggestions.\cite{BasleticPRB02,AudouardPRB94,VignollesPRB00}

\section{The new phase diagram of Bechgaard salts}

\begin{figure}
\includegraphics*[width=8cm]{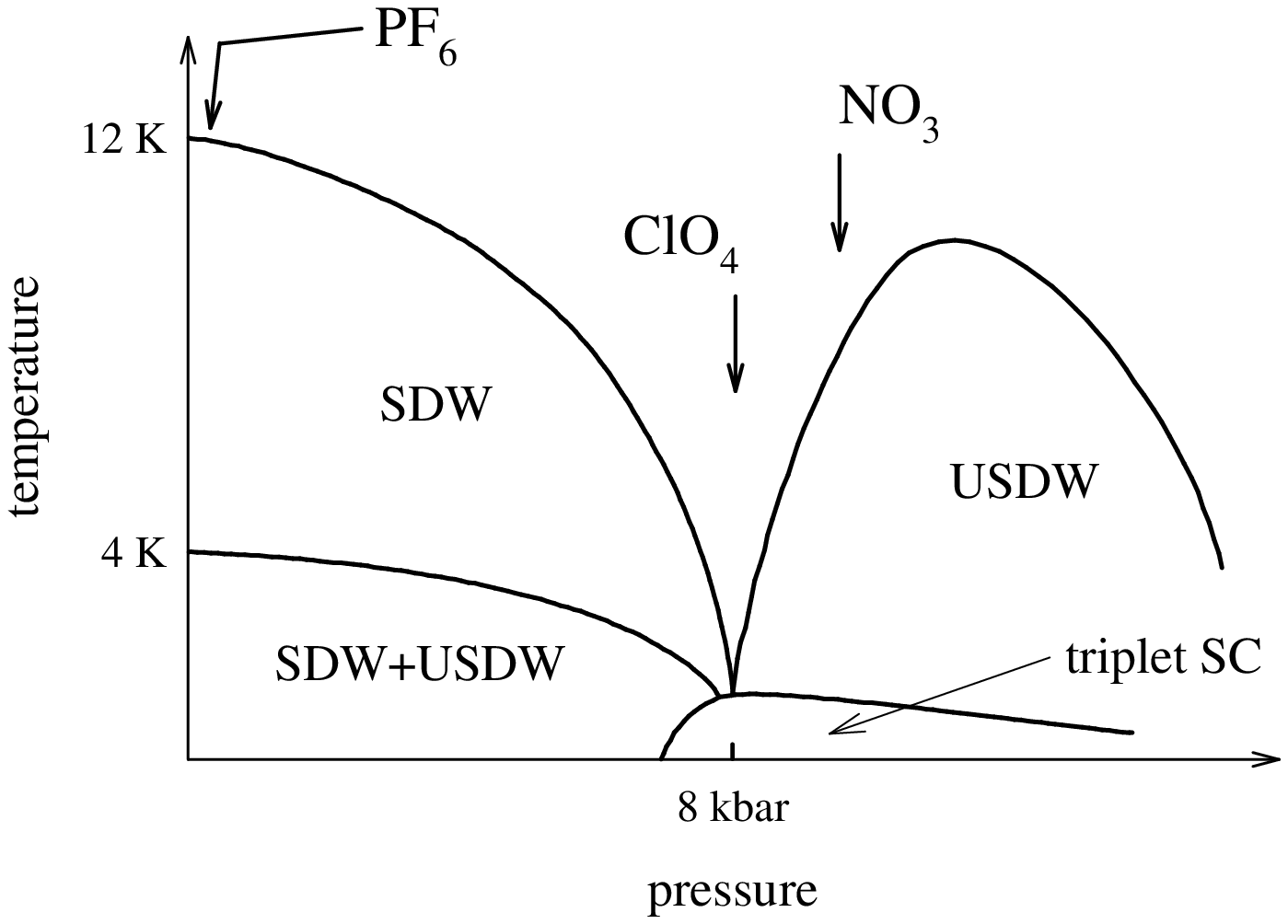}
\caption{\label{fig:phasediag} The schematic pressure--temperature
phase diagram for Bechgaard salts. Arrows denote position of \PF, \ClO\ and \NO\ in the phase diagram at ambient pressure.}
\end{figure}

Recently, one of us proposed phase diagram for Bechgaard salts with octahedral (centrosymmetric) anion like PF$_6$ which exhibit metallic behaviour down to the SDW transition at $\mTSDW\approx12\,$K (see Ref.\ \onlinecite{MakiCondMat0603806}). The salts with non-centrosymmetric anions undergo the AO transition and become insulating at ambient pressure, except for $X$$=$ClO$_4$ and NO$_3$. Here, we propose an extension/revision of the phase diagram (see \myreffig{fig:phasediag}). As indicated in \myreffig{fig:phasediag},
\PF\ at ambient pressure undergoes yet another transition around $T^\star\approx \mTSDW/3 \approx 4\,$K. The drastic change in the quasi-particle spectrum through $T^\star$ has been interpreted as appearance of SDW+USDW.\cite{BasleticPRB02} Further, from the angle dependent magnetoresistance of \PF\ and \ReO\ for $P > 8\,$kbar the existence of USDW in the high pressure range is inferred.\cite{KangSM03,DoraEPL04} Then, it is customarily to put \ClO\ at ambient pressure around $P = 8\,$kbar in \myreffig{fig:phasediag}, where transition from metallic state to superconducting one takes place. In this way we may understand the superconductivity at ambient pressure. Similarly, we may put \NO\ at ambient pressure around $P \simeq 8.5\,$kbar, since the transition from metallic state to density wave state takes place at $\mTSDW\approx9.5\,$K. The further behaviour of $\mTSDW$ vs.\ pressure is based on the experiments, which have shown that $\mTSDW$ is gradually suppressed under increasing pressure.\cite{KangPRL90} Then is the absence of superconductivity, and appearance of FISDW only at high pressure and high magnetic fields ($P \ge 8\,$kbar, $B > 20\,$T),\cite{VignollesPRB05} very surprising.

We think that the lack of inversion symmetry in NO$_3$ is at the heart of the absence of superconductivity and FISDW (for low pressure, $P < 8.5\,$kbar) in \NO. For example P.W.Anderson\cite{AndersonPRB84} speculated that the triplet superconductor cannot exist in a crystal without inversion symmetry. Also the nature of superconductivity in CePt$_3$Si, the crystal without inversion symmetry, is hotly discussed in the current literature.\cite{BauerPRL04,FrigeriNJP04} The inversion symmetry breaking is usually characterized by $E_{\text{ch}}$ the chiral symmetry breaking term or the Rashba term,\cite{FrigeriNJP04,RashbaSPSS60,GorkovPRL01} Both the absence of the triplet superconductivity (for $T_{\text{SC}} < 1\,$K) and the appearance of FISDW for $B>20\,$T suggest $2\,\text{K}<|E_{\text{ch}}|<10\,$K. Also this $E_{\text{ch}}$ appears to be consistent with $T_{\text{AO}}\approx45\,$K ($T_{\text{AO}}\gg E_{\text{ch}}$).\cite{WonPREPARATION} We believe that further study of the electronic properties of \NO\ is of great interest.

\section{Concluding Remarks}

We have shown that the anomalous low temperature behaviour of ADMR and Hall resistance of the Bechgaard salt \NO\ could be interpreted in terms of unconventional spin density wave, indicating that the possible ground state below $\mTSDW\approx 9.5\,$K is USDW. This is consistent with the new phase diagram of Bechgaard salts proposed recently,\cite{DoraEPL04,MakiCondMat0603806} which we -- in adition -- revised and extended in this paper. Therefore, it will be of great interest to study how the USDW order parameter $\Delta$ changes as the pressure is applied. This will provide a first step to explore the wider phase diagram. Another question is if there are other candidates in Bechgaard salts which exhibit USDW under ambient pressure.

We have also proposed possible explanation about the absence of superconductivity. Both the absence of superconductivity and
partial suppression of FISDW (i.e.\ the absence of FISDW for $P < 8\,$kbar, $B < 20\,$T) are due to the inversion symmetry breaking associated with the NO$_3$ anion ordering. The details on this will be published elsewhere.\cite{WonPREPARATION}

\begin{acknowledgments}
We thank Balazs D\'{o}ra, Peter Thalmeier and Hyekyung Won for
useful suggestions. K.M.\ thanks the hospitality of Max-Planck
Institut f\"{u}r Physik komplexer Systeme at Dresden, where the most
of this work is done.
\end{acknowledgments}


\begin{thebibliography}{34}
\expandafter\ifx\csname natexlab\endcsname\relax\def\natexlab#1{#1}\fi
\expandafter\ifx\csname bibnamefont\endcsname\relax
  \def\bibnamefont#1{#1}\fi
\expandafter\ifx\csname bibfnamefont\endcsname\relax
  \def\bibfnamefont#1{#1}\fi
\expandafter\ifx\csname citenamefont\endcsname\relax
  \def\citenamefont#1{#1}\fi
\expandafter\ifx\csname url\endcsname\relax
  \def\url#1{\texttt{#1}}\fi
\expandafter\ifx\csname urlprefix\endcsname\relax\def\urlprefix{URL }\fi
\providecommand{\bibinfo}[2]{#2}
\providecommand{\eprint}[2][]{\url{#2}}

\bibitem[{\citenamefont{Ishiguro et~al.}(1998)\citenamefont{Ishiguro, Yamaji,
  and Saito}}]{IshiguroBook98}
\bibinfo{author}{\bibfnamefont{T.}~\bibnamefont{Ishiguro}},
  \bibinfo{author}{\bibfnamefont{K.}~\bibnamefont{Yamaji}}, \bibnamefont{and}
  \bibinfo{author}{\bibfnamefont{G.}~\bibnamefont{Saito}},
  \emph{\bibinfo{title}{Organic superconductors}}
  (\bibinfo{publisher}{Springer}, \bibinfo{year}{1998}), \bibinfo{edition}{2nd}
  ed.

\bibitem[{\citenamefont{Mori}(2006)}]{MoriJPSJ06}
\bibinfo{author}{\bibfnamefont{H.}~\bibnamefont{Mori}}, \bibinfo{journal}{J.\
  Phys.\ Soc.\ Jpn.} \textbf{\bibinfo{volume}{75}}, \bibinfo{pages}{051003}
  (\bibinfo{year}{2006}).

\bibitem[{\citenamefont{{{Schegolev memorial volume}}}(1996)}]{Schegolev96}
\bibinfo{author}{\bibfnamefont{I.~F.} \bibnamefont{{{Schegolev memorial
  volume}}}}, \bibinfo{journal}{J.\ Phys. I (France)}
  \textbf{\bibinfo{volume}{6}}, \bibinfo{pages}{no.12} (\bibinfo{year}{1996}).

\bibitem[{\citenamefont{Lee et~al.}(2002)\citenamefont{Lee, Brown, Clark,
  Strouse, Naughton, Kang, and Chaikin}}]{LeePRL02}
\bibinfo{author}{\bibfnamefont{I.~J.} \bibnamefont{Lee}},
  \bibinfo{author}{\bibfnamefont{S.~E.} \bibnamefont{Brown}},
  \bibinfo{author}{\bibfnamefont{W.~G.} \bibnamefont{Clark}},
  \bibinfo{author}{\bibfnamefont{M.~J.} \bibnamefont{Strouse}},
  \bibinfo{author}{\bibfnamefont{M.~J.} \bibnamefont{Naughton}},
  \bibinfo{author}{\bibfnamefont{W.}~\bibnamefont{Kang}}, \bibnamefont{and}
  \bibinfo{author}{\bibfnamefont{P.~M.} \bibnamefont{Chaikin}},
  \bibinfo{journal}{Phys. Rev. Lett.} \textbf{\bibinfo{volume}{88}},
  \bibinfo{pages}{017004} (\bibinfo{year}{2002}).

\bibitem[{\citenamefont{Joo et~al.}(2005)\citenamefont{Joo, Auban-Senzier,
  Pasquier, J\'{e}rome, and Bechgaard}}]{JooEPL05}
\bibinfo{author}{\bibfnamefont{N.}~\bibnamefont{Joo}},
  \bibinfo{author}{\bibfnamefont{P.}~\bibnamefont{Auban-Senzier}},
  \bibinfo{author}{\bibfnamefont{C.~R.} \bibnamefont{Pasquier}},
  \bibinfo{author}{\bibfnamefont{D.}~\bibnamefont{J\'{e}rome}},
  \bibnamefont{and}
  \bibinfo{author}{\bibfnamefont{K.}~\bibnamefont{Bechgaard}},
  \bibinfo{journal}{Europhys.\ Lett.} \textbf{\bibinfo{volume}{72}},
  \bibinfo{pages}{645} (\bibinfo{year}{2005}).

\bibitem[{\citenamefont{Lee et~al.}(2006)\citenamefont{Lee, Brown, and
  Naughton}}]{LeeJPSJ06}
\bibinfo{author}{\bibfnamefont{I.~J.} \bibnamefont{Lee}},
  \bibinfo{author}{\bibfnamefont{S.~E.} \bibnamefont{Brown}}, \bibnamefont{and}
  \bibinfo{author}{\bibfnamefont{M.~J.} \bibnamefont{Naughton}},
  \bibinfo{journal}{J.\ Phys.\ Soc.\ Jpn.} \textbf{\bibinfo{volume}{75}},
  \bibinfo{pages}{051011} (\bibinfo{year}{2006}).

\bibitem[{\citenamefont{Kang et~al.}(2003)\citenamefont{Kang, Kang, Jo, and
  Uji}}]{KangSM03}
\bibinfo{author}{\bibfnamefont{W.}~\bibnamefont{Kang}},
  \bibinfo{author}{\bibfnamefont{H.}~\bibnamefont{Kang}},
  \bibinfo{author}{\bibfnamefont{Y.~J.} \bibnamefont{Jo}}, \bibnamefont{and}
  \bibinfo{author}{\bibfnamefont{S.}~\bibnamefont{Uji}},
  \bibinfo{journal}{Synth.\ Metals} \textbf{\bibinfo{volume}{133-134}},
  \bibinfo{pages}{15} (\bibinfo{year}{2003}).

\bibitem[{\citenamefont{D\'{o}ra
  et~al.}(2004{\natexlab{a}})\citenamefont{D\'{o}ra, Maki, Vanyolos, and
  Virosztek}}]{DoraEPL04}
\bibinfo{author}{\bibfnamefont{B.}~\bibnamefont{D\'{o}ra}},
  \bibinfo{author}{\bibfnamefont{K.}~\bibnamefont{Maki}},
  \bibinfo{author}{\bibfnamefont{A.}~\bibnamefont{Vanyolos}}, \bibnamefont{and}
  \bibinfo{author}{\bibfnamefont{A.}~\bibnamefont{Virosztek}},
  \bibinfo{journal}{Europhys.\ Lett.} \textbf{\bibinfo{volume}{67}},
  \bibinfo{pages}{1024} (\bibinfo{year}{2004}{\natexlab{a}}).

\bibitem[{\citenamefont{Tomi\'{c} et~al.}(1989)\citenamefont{Tomi\'{c}, Cooper,
  J\'{e}rome, and Bechgaard}}]{TomicPRL89}
\bibinfo{author}{\bibfnamefont{S.}~\bibnamefont{Tomi\'{c}}},
  \bibinfo{author}{\bibfnamefont{J.~R.} \bibnamefont{Cooper}},
  \bibinfo{author}{\bibfnamefont{D.}~\bibnamefont{J\'{e}rome}},
  \bibnamefont{and}
  \bibinfo{author}{\bibfnamefont{K.}~\bibnamefont{Bechgaard}},
  \bibinfo{journal}{Phys. Rev. Lett.} \textbf{\bibinfo{volume}{62}},
  \bibinfo{pages}{462} (\bibinfo{year}{1989}).

\bibitem[{\citenamefont{Yamaji}(1986)}]{YamajiSM86}
\bibinfo{author}{\bibfnamefont{K.}~\bibnamefont{Yamaji}},
  \bibinfo{journal}{Synth.\ Metals} \textbf{\bibinfo{volume}{13}},
  \bibinfo{pages}{29} (\bibinfo{year}{1986}).

\bibitem[{\citenamefont{Virosztek et~al.}(1986)\citenamefont{Virosztek, Chen,
  and Maki}}]{VirosztekPRB86}
\bibinfo{author}{\bibfnamefont{A.}~\bibnamefont{Virosztek}},
  \bibinfo{author}{\bibfnamefont{L.}~\bibnamefont{Chen}}, \bibnamefont{and}
  \bibinfo{author}{\bibfnamefont{K.}~\bibnamefont{Maki}},
  \bibinfo{journal}{Phys. Rev. B} \textbf{\bibinfo{volume}{34}},
  \bibinfo{pages}{3371} (\bibinfo{year}{1986}).

\bibitem[{\citenamefont{Poilblanc et~al.}(1986)\citenamefont{Poilblanc,
  H\'{e}ritier, Montambaux, and Lederer}}]{PoilblancJPC86}
\bibinfo{author}{\bibfnamefont{D.}~\bibnamefont{Poilblanc}},
  \bibinfo{author}{\bibfnamefont{M.}~\bibnamefont{H\'{e}ritier}},
  \bibinfo{author}{\bibfnamefont{G.}~\bibnamefont{Montambaux}},
  \bibnamefont{and} \bibinfo{author}{\bibfnamefont{P.}~\bibnamefont{Lederer}},
  \bibinfo{journal}{J. Phys. C} \textbf{\bibinfo{volume}{19}},
  \bibinfo{pages}{L321} (\bibinfo{year}{1986}).

\bibitem[{\citenamefont{Nersesyan and Vachnadze}(1989)}]{NersesyanJLTP89}
\bibinfo{author}{\bibfnamefont{A.~A.} \bibnamefont{Nersesyan}}
  \bibnamefont{and} \bibinfo{author}{\bibfnamefont{G.~E.}
  \bibnamefont{Vachnadze}}, \bibinfo{journal}{J.\ Low.\ Temp.\ Phys.}
  \textbf{\bibinfo{volume}{77}}, \bibinfo{pages}{293} (\bibinfo{year}{1989}).

\bibitem[{\citenamefont{Nersesyan et~al.}(1991)\citenamefont{Nersesyan,
  Japaridze, and Kimeridze}}]{NersesyanJPCM91}
\bibinfo{author}{\bibfnamefont{A.~A.} \bibnamefont{Nersesyan}},
  \bibinfo{author}{\bibfnamefont{G.~I.} \bibnamefont{Japaridze}},
  \bibnamefont{and} \bibinfo{author}{\bibfnamefont{I.~G.}
  \bibnamefont{Kimeridze}}, \bibinfo{journal}{J.\ Phys.\ Cond.\ Matt.}
  \textbf{\bibinfo{volume}{3}}, \bibinfo{pages}{3353} (\bibinfo{year}{1991}).

\bibitem[{\citenamefont{D\'{o}ra
  et~al.}(2004{\natexlab{b}})\citenamefont{D\'{o}ra, Maki, and
  Virosztek}}]{DoraMPL04}
\bibinfo{author}{\bibfnamefont{B.}~\bibnamefont{D\'{o}ra}},
  \bibinfo{author}{\bibfnamefont{K.}~\bibnamefont{Maki}}, \bibnamefont{and}
  \bibinfo{author}{\bibfnamefont{A.}~\bibnamefont{Virosztek}},
  \bibinfo{journal}{Mod. Phys. Lett. B} \textbf{\bibinfo{volume}{18}},
  \bibinfo{pages}{327} (\bibinfo{year}{2004}{\natexlab{b}}).

\bibitem[{\citenamefont{Maki et~al.}()\citenamefont{Maki, D\'{o}ra, and
  Virosztek}}]{MakiCondMat0603806}
\bibinfo{author}{\bibfnamefont{K.}~\bibnamefont{Maki}},
  \bibinfo{author}{\bibfnamefont{B.}~\bibnamefont{D\'{o}ra}}, \bibnamefont{and}
  \bibinfo{author}{\bibfnamefont{A.}~\bibnamefont{Virosztek}},
  \eprint{cond-mat/0603806}.

\bibitem[{\citenamefont{Bi\v{s}kup et~al.}(1994)\citenamefont{Bi\v{s}kup,
  Balicas, Tomi\'{c}, J\'{e}rome, and Fabre}}]{BiskupPRB94}
\bibinfo{author}{\bibfnamefont{N.}~\bibnamefont{Bi\v{s}kup}},
  \bibinfo{author}{\bibfnamefont{L.}~\bibnamefont{Balicas}},
  \bibinfo{author}{\bibfnamefont{S.}~\bibnamefont{Tomi\'{c}}},
  \bibinfo{author}{\bibfnamefont{D.}~\bibnamefont{J\'{e}rome}},
  \bibnamefont{and} \bibinfo{author}{\bibfnamefont{J.~M.} \bibnamefont{Fabre}},
  \bibinfo{journal}{Phys. Rev. B} \textbf{\bibinfo{volume}{50}},
  \bibinfo{pages}{12721} (\bibinfo{year}{1994}).

\bibitem[{\citenamefont{Basleti\'{c} et~al.}(1996)\citenamefont{Basleti\'{c},
  Korin-Hamzi\'{c}, Hamzi\'{c}, Tomi\'{c}, and Fabre}}]{BasleticSSC96}
\bibinfo{author}{\bibfnamefont{M.}~\bibnamefont{Basleti\'{c}}},
  \bibinfo{author}{\bibfnamefont{B.}~\bibnamefont{Korin-Hamzi\'{c}}},
  \bibinfo{author}{\bibfnamefont{A.}~\bibnamefont{Hamzi\'{c}}},
  \bibinfo{author}{\bibfnamefont{S.}~\bibnamefont{Tomi\'{c}}},
  \bibnamefont{and} \bibinfo{author}{\bibfnamefont{J.~M.} \bibnamefont{Fabre}},
  \bibinfo{journal}{Solid State Commun.} \textbf{\bibinfo{volume}{97}},
  \bibinfo{pages}{333} (\bibinfo{year}{1996}).

\bibitem[{\citenamefont{Bi\v{s}kup et~al.}(1993)\citenamefont{Bi\v{s}kup,
  Basleti\'{c}, Tomi\'{c}, Korin-Hamzi\'{c}, Maki, Bechgaard, and
  Fabre}}]{BiskupPRB93}
\bibinfo{author}{\bibfnamefont{N.}~\bibnamefont{Bi\v{s}kup}},
  \bibinfo{author}{\bibfnamefont{M.}~\bibnamefont{Basleti\'{c}}},
  \bibinfo{author}{\bibfnamefont{S.}~\bibnamefont{Tomi\'{c}}},
  \bibinfo{author}{\bibfnamefont{B.}~\bibnamefont{Korin-Hamzi\'{c}}},
  \bibinfo{author}{\bibfnamefont{K.}~\bibnamefont{Maki}},
  \bibinfo{author}{\bibfnamefont{K.}~\bibnamefont{Bechgaard}},
  \bibnamefont{and} \bibinfo{author}{\bibfnamefont{J.~M.} \bibnamefont{Fabre}},
  \bibinfo{journal}{Phys. Rev. B} \textbf{\bibinfo{volume}{47}},
  \bibinfo{pages}{8289} (\bibinfo{year}{1993}).

\bibitem[{\citenamefont{Basleti\'{c} et~al.}(2002)\citenamefont{Basleti\'{c},
  Korin-Hamzi\'{c}, and Maki}}]{BasleticPRB02}
\bibinfo{author}{\bibfnamefont{M.}~\bibnamefont{Basleti\'{c}}},
  \bibinfo{author}{\bibfnamefont{B.}~\bibnamefont{Korin-Hamzi\'{c}}},
  \bibnamefont{and} \bibinfo{author}{\bibfnamefont{K.}~\bibnamefont{Maki}},
  \bibinfo{journal}{Phys. Rev. B} \textbf{\bibinfo{volume}{65}},
  \bibinfo{pages}{235117} (\bibinfo{year}{2002}).

\bibitem[{\citenamefont{D\'{o}ra et~al.}(2002)\citenamefont{D\'{o}ra, Maki,
  Korin-Hamzi\'{c}, Basleti\'{c}, Virosztek, Kartsovnik, and
  M{\"{u}}ller}}]{DoraEPL02}
\bibinfo{author}{\bibfnamefont{B.}~\bibnamefont{D\'{o}ra}},
  \bibinfo{author}{\bibfnamefont{K.}~\bibnamefont{Maki}},
  \bibinfo{author}{\bibfnamefont{B.}~\bibnamefont{Korin-Hamzi\'{c}}},
  \bibinfo{author}{\bibfnamefont{M.}~\bibnamefont{Basleti\'{c}}},
  \bibinfo{author}{\bibfnamefont{A.}~\bibnamefont{Virosztek}},
  \bibinfo{author}{\bibfnamefont{M.~V.} \bibnamefont{Kartsovnik}},
  \bibnamefont{and}
  \bibinfo{author}{\bibfnamefont{H.}~\bibnamefont{M{\"{u}}ller}},
  \bibinfo{journal}{Europhys.\ Lett.} \textbf{\bibinfo{volume}{60}},
  \bibinfo{pages}{737} (\bibinfo{year}{2002}).

\bibitem[{\citenamefont{Maki et~al.}(2003)\citenamefont{Maki, D\'{o}ra,
  Kartsovnik, Virosztek, Korin-Hamzi\'{c}, and M.Basleti\'{c}}}]{MakiPRL03}
\bibinfo{author}{\bibfnamefont{K.}~\bibnamefont{Maki}},
  \bibinfo{author}{\bibfnamefont{B.}~\bibnamefont{D\'{o}ra}},
  \bibinfo{author}{\bibfnamefont{M.}~\bibnamefont{Kartsovnik}},
  \bibinfo{author}{\bibfnamefont{A.}~\bibnamefont{Virosztek}},
  \bibinfo{author}{\bibfnamefont{B.}~\bibnamefont{Korin-Hamzi\'{c}}},
  \bibnamefont{and} \bibinfo{author}{\bibnamefont{M.Basleti\'{c}}},
  \bibinfo{journal}{Phys. Rev. Lett.} \textbf{\bibinfo{volume}{90}},
  \bibinfo{pages}{256402} (\bibinfo{year}{2003}).

\bibitem[{\citenamefont{Nambu}(1960)}]{NambuPR60}
\bibinfo{author}{\bibfnamefont{Y.}~\bibnamefont{Nambu}},
  \bibinfo{journal}{Phys. Rev.} \textbf{\bibinfo{volume}{117}},
  \bibinfo{pages}{648} (\bibinfo{year}{1960}).

\bibitem[{\citenamefont{Basleti\'{c} et~al.}(1993)\citenamefont{Basleti\'{c},
  Bi\v{s}kup, Korin-Hamzi\'{c}, Tomi\'{c}, Hamzi\'{c}, Bechgaard, and
  Fabre}}]{BasleticEPL93}
\bibinfo{author}{\bibfnamefont{M.}~\bibnamefont{Basleti\'{c}}},
  \bibinfo{author}{\bibfnamefont{N.}~\bibnamefont{Bi\v{s}kup}},
  \bibinfo{author}{\bibfnamefont{B.}~\bibnamefont{Korin-Hamzi\'{c}}},
  \bibinfo{author}{\bibfnamefont{S.}~\bibnamefont{Tomi\'{c}}},
  \bibinfo{author}{\bibfnamefont{A.}~\bibnamefont{Hamzi\'{c}}},
  \bibinfo{author}{\bibfnamefont{K.}~\bibnamefont{Bechgaard}},
  \bibnamefont{and} \bibinfo{author}{\bibfnamefont{J.~M.} \bibnamefont{Fabre}},
  \bibinfo{journal}{Europhys.\ Lett.} \textbf{\bibinfo{volume}{22}},
  \bibinfo{pages}{279} (\bibinfo{year}{1993}).

\bibitem[{\citenamefont{Audouard et~al.}(1994)\citenamefont{Audouard, Goze,
  Ulmet, Brossard, Askenazy, and Fabre}}]{AudouardPRB94}
\bibinfo{author}{\bibfnamefont{A.}~\bibnamefont{Audouard}},
  \bibinfo{author}{\bibfnamefont{F.}~\bibnamefont{Goze}},
  \bibinfo{author}{\bibfnamefont{J.-P.} \bibnamefont{Ulmet}},
  \bibinfo{author}{\bibfnamefont{L.}~\bibnamefont{Brossard}},
  \bibinfo{author}{\bibfnamefont{S.}~\bibnamefont{Askenazy}}, \bibnamefont{and}
  \bibinfo{author}{\bibfnamefont{J.-M.} \bibnamefont{Fabre}},
  \bibinfo{journal}{Phys. Rev. B} \textbf{\bibinfo{volume}{50}},
  \bibinfo{pages}{12726} (\bibinfo{year}{1994}).

\bibitem[{\citenamefont{Vignolles et~al.}(2000)\citenamefont{Vignolles, Ulmet,
  Audouard, Naughton, and Fabre}}]{VignollesPRB00}
\bibinfo{author}{\bibfnamefont{D.}~\bibnamefont{Vignolles}},
  \bibinfo{author}{\bibfnamefont{J.~P.} \bibnamefont{Ulmet}},
  \bibinfo{author}{\bibfnamefont{A.}~\bibnamefont{Audouard}},
  \bibinfo{author}{\bibfnamefont{M.~J.} \bibnamefont{Naughton}},
  \bibnamefont{and} \bibinfo{author}{\bibfnamefont{J.~M.} \bibnamefont{Fabre}},
  \bibinfo{journal}{Phys. Rev. B} \textbf{\bibinfo{volume}{61}},
  \bibinfo{pages}{8913} (\bibinfo{year}{2000}).

\bibitem[{\citenamefont{Kang et~al.}(1990)\citenamefont{Kang, Hannahs, Chiang,
  Upasani, and Chaikin}}]{KangPRL90}
\bibinfo{author}{\bibfnamefont{W.}~\bibnamefont{Kang}},
  \bibinfo{author}{\bibfnamefont{S.~T.} \bibnamefont{Hannahs}},
  \bibinfo{author}{\bibfnamefont{L.~Y.} \bibnamefont{Chiang}},
  \bibinfo{author}{\bibfnamefont{R.}~\bibnamefont{Upasani}}, \bibnamefont{and}
  \bibinfo{author}{\bibfnamefont{P.~M.} \bibnamefont{Chaikin}},
  \bibinfo{journal}{Phys. Rev. Lett.} \textbf{\bibinfo{volume}{65}},
  \bibinfo{pages}{2812} (\bibinfo{year}{1990}).

\bibitem[{\citenamefont{Vignolles et~al.}(2005)\citenamefont{Vignolles,
  Audouard, Nardone, Brossard, Bouguessa, and Fabre}}]{VignollesPRB05}
\bibinfo{author}{\bibfnamefont{D.}~\bibnamefont{Vignolles}},
  \bibinfo{author}{\bibfnamefont{A.}~\bibnamefont{Audouard}},
  \bibinfo{author}{\bibfnamefont{M.}~\bibnamefont{Nardone}},
  \bibinfo{author}{\bibfnamefont{L.}~\bibnamefont{Brossard}},
  \bibinfo{author}{\bibfnamefont{S.}~\bibnamefont{Bouguessa}},
  \bibnamefont{and} \bibinfo{author}{\bibfnamefont{J.-M.} \bibnamefont{Fabre}},
  \bibinfo{journal}{Phys. Rev. B} \textbf{\bibinfo{volume}{71}},
  \bibinfo{pages}{020404(R)} (\bibinfo{year}{2005}).

\bibitem[{\citenamefont{Anderson}(1984)}]{AndersonPRB84}
\bibinfo{author}{\bibfnamefont{P.~W.} \bibnamefont{Anderson}},
  \bibinfo{journal}{Phys. Rev. B} \textbf{\bibinfo{volume}{30}},
  \bibinfo{pages}{4000} (\bibinfo{year}{1984}).

\bibitem[{\citenamefont{Bauer et~al.}(2004)\citenamefont{Bauer, Hilscher,
  Michor, Paul, Scheidt, Gribanov, Seropegin, No{\"{e}}l, Sigrist, and
  Rogl}}]{BauerPRL04}
\bibinfo{author}{\bibfnamefont{E.}~\bibnamefont{Bauer}},
  \bibinfo{author}{\bibfnamefont{G.}~\bibnamefont{Hilscher}},
  \bibinfo{author}{\bibfnamefont{H.}~\bibnamefont{Michor}},
  \bibinfo{author}{\bibfnamefont{C.}~\bibnamefont{Paul}},
  \bibinfo{author}{\bibfnamefont{E.~W.} \bibnamefont{Scheidt}},
  \bibinfo{author}{\bibfnamefont{A.}~\bibnamefont{Gribanov}},
  \bibinfo{author}{\bibfnamefont{Y.}~\bibnamefont{Seropegin}},
  \bibinfo{author}{\bibfnamefont{H.}~\bibnamefont{No{\"{e}}l}},
  \bibinfo{author}{\bibfnamefont{M.}~\bibnamefont{Sigrist}}, \bibnamefont{and}
  \bibinfo{author}{\bibfnamefont{P.}~\bibnamefont{Rogl}},
  \bibinfo{journal}{Phys. Rev. Lett.} \textbf{\bibinfo{volume}{92}},
  \bibinfo{pages}{027003} (\bibinfo{year}{2004}).

\bibitem[{\citenamefont{Frigeri et~al.}(2004)\citenamefont{Frigeri, Agterberg,
  and Sigrist}}]{FrigeriNJP04}
\bibinfo{author}{\bibfnamefont{P.~A.} \bibnamefont{Frigeri}},
  \bibinfo{author}{\bibfnamefont{D.~F.} \bibnamefont{Agterberg}},
  \bibnamefont{and} \bibinfo{author}{\bibfnamefont{M.}~\bibnamefont{Sigrist}},
  \bibinfo{journal}{New J. Physics} \textbf{\bibinfo{volume}{6}},
  \bibinfo{pages}{115} (\bibinfo{year}{2004}).

\bibitem[{\citenamefont{Rashba}(1960)}]{RashbaSPSS60}
\bibinfo{author}{\bibfnamefont{E.~I.} \bibnamefont{Rashba}},
  \bibinfo{journal}{Sov. Phys. Solid State} \textbf{\bibinfo{volume}{2}},
  \bibinfo{pages}{1709} (\bibinfo{year}{1960}).

\bibitem[{\citenamefont{Gor'kov and Rashba}(2001)}]{GorkovPRL01}
\bibinfo{author}{\bibfnamefont{L.~P.} \bibnamefont{Gor'kov}} \bibnamefont{and}
  \bibinfo{author}{\bibfnamefont{E.~I.} \bibnamefont{Rashba}},
  \bibinfo{journal}{Phys. Rev. Lett.} \textbf{\bibinfo{volume}{87}},
  \bibinfo{pages}{037004} (\bibinfo{year}{2001}).

\bibitem[{\citenamefont{Won and Maki}()}]{WonPREPARATION}
\bibinfo{author}{\bibfnamefont{H.}~\bibnamefont{Won}} \bibnamefont{and}
  \bibinfo{author}{\bibfnamefont{K.}~\bibnamefont{Maki}}, \bibinfo{note}{in
  preparation}.

\end{thebibliography}
\end{document}